


\input phyzzx
\newdimen\tableauside\tableauside=1.0ex
\newdimen\tableaurule\tableaurule=0.4pt
\newdimen\tableaustep
\def\phantomhrule#1{\hbox{\vbox
to0pt{\hrule height\tableaurule width#1\vss}}}
\def\phantomvrule#1{\vbox{\hbox
to0pt{\vrule width\tableaurule height#1\hss}}}
\def\sqr{\vbox{%
\phantomhrule\tableaustep
\hbox{\phantomvrule\tableaustep\kern\tableaustep\phantomvrule\tableaustep}%
\hbox{\vbox{\phantomhrule\tableauside}\kern-\tableaurule}}}
\def\squares#1{\hbox{\count0=#1\noindent\loop\sqr
  \advance\count0 by-1 \ifnum\count0>0\repeat}}
\def\tableau#1{\vcenter{\offinterlineskip
  \tableaustep=\tableauside\advance\tableaustep by-\tableaurule
  \kern\normallineskip\hbox
    {\kern\normallineskip\vbox
      {\gettableau#1 0 }%
     \kern\normallineskip\kern\tableaurule}%
  \kern\normallineskip\kern\tableaurule}}
\def\gettableau#1 {\ifnum#1=0\let\next=\null\else
  \squares{#1}\let\next=\gettableau\fi\next}

\tableauside=1.8ex
\tableaurule=0.4pt

\def\np{Nucl. Phys.}
\def\pl{Phys. Lett.}

\def\pr{Phys. Rev.}

\def\cmp{Comm. Math. Phys.}

\def\mpl{Mod. Phys. Lett.}
\def\half{{1\over 2}}

\def\tr{{\hbox{\rm Tr}}}

\def\ele{{\hbox{\sevenrm L}}}
\def\ere{{\hbox{\sevenrm R}}}
\def\lai{{\hbox{\sevenrm I}}}
\def\laf{{\hbox{\sevenrm f}}}
\def\laga{{\hbox{\sevenrm e,A}}}

\def\D{\nabla}

\tolerance=500000
\overfullrule=0pt

\Pubnum={US-FT-2/93\cr
hep-th/9305028\cr
revised version}
\pubnum={US-FT-2/93}
\date={April, 1993}
\pubtype={}
\titlepage

\title{BREAKING OF TOPOLOGICAL SYMMETRY}
 \author{ M. Alvarez
and J.M.F. Labastida\foot{e-mail: LABASTIDA@GAES.USC.ES} }
\address{Departamento de F\'\i sica de Part\'\i culas\break Universidade de
Santiago\break E-15706 Santiago de Compostela, Spain}

\abstract{The coupling of topological matter to topological
Yang-Mills theory in four dimensions is considered and a model is
presented. It is shown that, contrary to the two-dimensional case, this
coupling may lead to a breaking of the topological symmetry. This means that
the vacuum expectation values of the observables of the theory loose their
invariance under small deformations of the metric  while
the action of  the model  possesses all the symmetries
corresponding to the case with no coupling.}

\endpage
\pagenumber=1
\sequentialequations

Matter couplings to topological Yang-Mills and topological gravity in
two dimensions can be constructed without loosing the topological
features of the theory \REF\tsm{E. Witten\journal\cmp&118(88)411}
\REF\tg{J.M.F. Labastida, M. Pernici and E.
Witten\journal\np&310(88)611} \REF\pablo{J.M.F. Labastida and P.M.
Llatas\journal\np&379(92)220} [\tsm,\tg,\pablo]. In other
words, these couplings can be  constructed while maintaining the
$Q$-symmetry of the models, and it turns out that the resulting theory
possesses an energy-momentum tensor which is $Q$-exact. The aim of
this letter is to point out that in four dimensions the picture that
emerges seems to be  different.  We will present a simple model of
topological matter in four dimensions and its coupling to topological
Yang-Mills theory (or Donaldson-Witten theory) \REF\tqft{E.
Witten\journal\cmp&117(88)353} [\tqft]. The resulting theory has a
$Q$-symmetry but its energy-momentum tensor is not $Q$-exact. This
implies that the observables leading to Donaldson invariants in
topological Yang-Mills theory may get non-topological corrections due to
the presence of matter couplings. We will show that it is also
possible to add mass terms while preserving $Q$-invariance, leading to
a further breaking of the topological character of the theory.

Let us begin constructing topological matter in four dimensions. Our
starting point is a twisted version of the hypermultiplet of $N=2$
supersymmetry \REF\fayet{P. Fayet\journal\np&B113(76)135}
\REF\sohn{M.F. Sohnius\journal\np&B138(78)109} [\fayet,\sohn].
The resulting  models are different than the ones obtained after twisting
$N=4$ supersymmetry, or $N=2$ conformal supergravity as constructed in
\REF\roc{A. Karlhede and M. Ro\v cek\journal\pl&B212(88)51}
\REF\yam{J. Yamron\journal\pl&B213(88)325} [\roc] and [\yam],
respectively. In four dimensions, the Lorentz and internal generators
of $N=2$ supersymmetry can be grouped as the ones of
 $SU(2)_\ele\times SU(2)_\ere\times
SU(2)_\lai \times U(1)$.
The hypermultiplet is made out of four fields which transform as
$(0,0,1/2)^0$, $(1/2,0,0)^1$, $(0,1/2,0)^{-1}$ and $(0,0,1/2)^2$
respect to those generators. The superindex denotes the $U(1)$
eigenvalue. The field transforming as $(0,0,1/2)^2$ is auxiliary. The
twisting  consists of the replacement of $SU(2)_\ere \times SU(2)_\lai$
by  $SU(2)_\ere'$, being this the diagonal sum of $SU(2)_\ere$ and
$SU(2)_\lai$. Under the twisted algebra $SU(2)_\ele\times SU(2)_\ere'
\times U(1)$ the component fields of the twisted hypermultiplet
transform as $(0,1/2)^0$, $(0,1/2)^1$, $(1/2,0)^{-1}$ and $(0,1/2)^2$.
We will denote these fields as $H_\alpha$, $u_\alpha$,
$v_{\dot\alpha}$ and $K_\alpha$ respectively. Clearly, $H_\alpha$ and
$K_\alpha$ are commuting fields while  $u_\alpha$ and
$v_{\dot\alpha}$ are anticommuting. The $U(1)$ quantum numbers of the
$N=2$ fields now play the role of ghost numbers. The resulting
$Q$-transformations are,
$$ \eqalign{
\delta H_{\alpha} &= \epsilon u_{\alpha}, \cr
\delta u_{\alpha} &= -\epsilon K_{\alpha}, \cr
\delta v_{\dot\alpha}
&=i\epsilon\partial_{{\alpha\dot\alpha}}H^{\alpha},\cr   \delta K_{\alpha}
&=i\epsilon\partial_{{\alpha\dot\alpha}} v^{\dot\alpha}, \cr}
\eqn\castana
$$
where $\epsilon$ is a constant anticommuting parameter. These
transformations indicate that $Q^2\neq 0$. This is not surprising
since the $N=2$ hypermultiplet possesses central charges. Other models
of topological quantum field theories where central charges are
present have been previously studied in two dimensions
\REF\pot{J.M.F. Labastida and P.M. Llatas\journal\pl&B271(91)101}
[\pot,\pablo]. The commuting central charge transformations can be
easily found from \castana\ and the fact that $2Q^2 = Z$, where $Z$ is
the central charge generator. They turn out to be,
$$
\eqalign{
\delta_{z} H_{\alpha} &= - z K_{\alpha}, \cr
\delta_{z} u_{\alpha} &= -i z
\partial_{\alpha\dot\alpha} v^{\dot\alpha},
\cr \delta_{z} v_{\dot\alpha}
&=i z\partial_{\alpha\dot\alpha}u^{\alpha},\cr    \delta_{z} K_{\alpha}
&= z \tableau{1} H_\alpha. \cr}   \eqn\castanados
$$
where $z$ is a commuting constant parameter. It is simple to
verify that indeed,
$$
[Q,Z] = 0.
\eqn\raul
$$
The presence of central charges breaks the $U(1)$ symmetry, which in
the twisted theory is just the ghost number symmetry, into $Z_4$. It
can be verified explicitly that, indeed, the ghost number is preserved
in the transformations \castana\ and \castanados\ modulo 4.

To construct matter actions we will introduce a second multiplet
which can be thought as the complex conjugate of the one just
described. We will denote the component fields of this multiplet by
$\overline{H}_\alpha$, $\overline{u}_\alpha$,
$\overline{v}_{\dot\alpha}$ and $\overline{K}_\alpha$. Clearly, under
$Q$ and $Z$ they have the same  transformation properties as in
\castana\ and \castanados.  For example, the corresponding $Q$
transformations are,
$$ \eqalign{
\delta \overline{H}_{\alpha} &= \epsilon \overline{u}_{\alpha}, \cr
\delta \overline{u}_{\alpha} &= - \epsilon
\overline{K}_{\alpha}, \cr
\delta \overline{v}_{\dot\alpha}
&=i\epsilon\partial_{{\alpha\dot\alpha}}\overline{H}^{\alpha},\cr   \delta
\overline{K}_{\alpha} &=i\epsilon\partial_{{\alpha\dot\alpha}}
\overline{v}^{\dot\alpha}, \cr}  \eqn\castanon
$$
The ghost number assignment for these fields is the same as the one
used for their counterparts with no overlines. The matter action which
is invariant under $Q$ and $Z$ takes the form:
$$
\eqalign{
{\cal L}^\laf = & {\cal L}^\laf_0 + m{\cal L}^\laf_m \cr
  = &\int d^4x \Big[
\overline{H}^{\alpha}\tableau{1} H_{\alpha}+{i}
\overline{u}^{\alpha}\partial_{{\alpha\dot\alpha}}{v}^{\dot\alpha} -
{i}\overline{v}^{\dot\alpha}\partial_{{\alpha\dot\alpha}}u^{\alpha}
+\overline{K}^{\alpha}K_{\alpha}
\cr
&\,\,\,\,\,\,\,\,\,\,\,\,\,\,\,\,\,\,\,
+m \big(\overline{K}^{\alpha}H_{\alpha}
- \overline{H}^{\alpha}K_{\alpha}\big)
+ m(\overline{u}^{\alpha}u_\alpha +
\overline{v}^{\dot\alpha}v_{\dot\alpha})\Big],\cr}
\eqn\nuezmoscada
$$
where $m$ is the bare mass associated to the twisted hypermultiplet.
Notice that $K_\alpha$ and $\overline{K}_\alpha$ play the role of
auxiliary fields. For $m\neq 0$ it is convenient to redefine these
fields so that they appear cuadratically in the action:
$$
\eqalign{\overline{K}_{\alpha} &=\overline{K'}_{\alpha}+
m\overline{H}_{\alpha}, \cr
K_{\alpha} &= {K'}_{\alpha}- m H_{\alpha}. \cr}
\eqn\Recesvinto
$$
The mass terms in \nuezmoscada\ become,
$$
m{\cal L}^\laf_m = \int d^4x  \big[
m^2 \overline{H}^\alpha H_\alpha
+m(\overline{u}^\alpha u_\alpha +
\overline{v}^{\dot\alpha}v_{\dot\alpha})
\big].
\eqn\nuezmoscadados
$$

In order to study if this model leads to a topological quantum field
theory we must formulate it for an arbitrary four-manifold. Since the
model contains spinors this manifold must admit at least one spin
structure, \ie, it must be a spin manifold. Actually, we will endow
our manifold also with an $SU(N)$ gauge connection $A_\mu$. Vector
indices will be denoted by greek letters from the middle of the
alphabet. The ones from the beginning of the alphabet are reserved for
spinor indices. Let us denote by ${M}$ an arbitrary four-dimensional
spin manifold without boundary. Introducing a vierbein $e_{a\mu}$ on
such a manifold, one finds that the action
$$
\eqalign{
{\cal L}^\laga = & {\cal L}^\laga_0 + m{\cal L}^\laga_m \cr
  = &\int_{M} d^4x e \Big[
\overline{H}^{\alpha}(\tableau{1}+{1\over 4}R) H_{\alpha}
+{i\over 2}\overline{H}^\alpha F_{\alpha\beta}^{+} H^\beta +{i}
\overline{u}^{\alpha}\nabla_{{\alpha\dot\alpha}}{v}^{\dot\alpha} -
{i}\overline{v}^{\dot\alpha}\nabla_{{\alpha\dot\alpha}}u^{\alpha}
\cr
&\,\,\,\,\,\,\,\,\,\,\,\,\,\,\,\,\,\,\,\,\,\,\,\,\,\,\,\,\,\,
+\overline{K'}^{\alpha}K_{\alpha}'
+ m^2 \overline{H}^\alpha H_\alpha
+m(\overline{u}^\alpha u_\alpha +
\overline{v}^{\dot\alpha}v_{\dot\alpha})\Big],\cr}
\eqn\maria
$$
is invariant under the following $Q$ and $Z$ transformations,
$$
\eqalign{
\delta H_{\alpha} &= \epsilon u_{\alpha}, \cr
\delta u_{\alpha} &= -\epsilon K_{\alpha}, \cr
\delta v_{\dot\alpha}
&=i\epsilon\nabla_{{\alpha\dot\alpha}}H^{\alpha},\cr   \delta K_{\alpha}
&=i\epsilon\nabla_{{\alpha\dot\alpha}} v^{\dot\alpha}, \cr}
\qquad\qquad
\eqalign{
\delta_{z} H_{\alpha} &= - z K_{\alpha}, \cr
\delta_{z} u_{\alpha} &= -iz
\nabla_{\alpha\dot\alpha} v^{\dot\alpha},
\cr \delta_{z} v_{\dot\alpha} &=iz
\nabla_{\alpha\dot\alpha}u^{\alpha},\cr
\delta_{z} K_{\alpha} &= z \big[(\tableau{1}+{1\over 4}R) H_\alpha
+{i\over 2} F^{+}_{\alpha\beta}H^\beta\big]. \cr}
\eqn\yago
$$
In \maria\ and \yago\  $\nabla_\mu$ denotes a covariant derivative
respect to the vierbein $e_{a\mu}$ and the gauge connection $A_\mu$,
$R$ is the curvature scalar, and $F^{+}_{\alpha\beta}$ is the
anti-self-dual part of the gauge field strength, $F^{+}_{\alpha\beta}=
C^{\dot\alpha\dot\beta}F_{\alpha\dot\alpha,\beta\dot\beta}$. Of course,
$\tableau{1}$ denotes the covariant Laplacian. The matter fields with
no overline transform under a given representation of $SU(N)$ while
the ones with overlines transform under the complex conjugate
representation.  Notice that in this generalized setting the
transformations \yago\ get  terms involving the Riemann
curvature and the gauge field strength. A similar set of
transformations as the ones in \yago\ holds for the fields with
overline. The generalized transformations \yago\ verify the algebra
\raul.

Certainly, the action ${\cal L}^\laga$ in \maria\ is not topological
because of the mass terms. However, if $m=0$ the action is
topological. To verify this we must analyze whether or not the
energy-momentum tensor is $Q$-exact. First, notice that the action is
$Q$-exact. It turns out that,
$$
{\cal L}^\laga_0 = \big\{ Q, \Lambda^\laga  \big\},
\eqn\adri
$$
where
$$
\Lambda^\laga =\half \int_{M} d^4x e \Big[
i\overline{H}^\alpha \nabla_{\alpha\dot\alpha} v^{\dot\alpha}
+i\overline{v}^{\dot\alpha}\D_{\alpha\dot\alpha}H^{\alpha}
-\overline{K}^\alpha u_\alpha
-\overline{u}^{\alpha}K_{\alpha}\Big].
\eqn\adriana
$$
The invariance under $Q$ and $Z$ of ${\cal L}^\laga_0$ follows simply
from  \raul\ and the fact that,
$$
[Z,\Lambda]=0.
\eqn\laura
$$
The form of \adri\ does not imply in general that the theory is topological.
Only when $Q$ and ${\delta\over \delta e^{a\mu}}$  commute this implication
holds. This is not the case, however, in this model as can be concluded
from the transformations \yago. Indeed, it turns out that the
energy-momentum tensor is not $Q$-exact. One finds:
$$
T_{\mu\nu}^\laga =  \big\{ Q , \Lambda_{\mu\nu}^\laga \big\}
+ {1\over 2}g_{\mu\nu}  T^\laga_0,
\eqn\newuno
$$
where,
$$
\eqalign{
\Lambda_{\mu\nu}^\laga = &{i\over 4}\big[
\overline H^\alpha
(\sigma_{(\mu})_{\alpha\dot\beta} \nabla_{\nu)} v^{\dot\beta}
-\nabla_{(\mu} \overline v^{\dot\beta} (\sigma_{\nu)})_{\alpha\dot\beta}
H^\alpha \cr
&\,\,\,\,\,-\nabla_{(\mu} \overline H^{\alpha}
(\sigma_{\nu)})_{\alpha\dot\beta} v^{\dot\beta}
+\overline v^{\dot\beta}
(\sigma_{(\mu})_{\alpha\dot\beta} \nabla_{\nu)} H^{\alpha}\big] \cr
-&{i\over 2}\big[
\overline H^\alpha
(\sigma_{\mu\nu})_{\alpha}{}^{\beta} \nabla_{\beta\dot\beta} v^{\dot\beta}
-\nabla^{\gamma}{}_{\dot\alpha} \overline v^{\dot\alpha}
(\sigma_{\mu\nu})_{\gamma}{}^{\alpha} H^\alpha \cr
&\,\,\,\,\,+\nabla_{\alpha\dot\beta} \overline H^{\alpha}
(\tilde\sigma_{\mu\nu})^{\dot\beta}{}_{\dot\alpha} v^{\dot\alpha}
-\overline v_{\dot\alpha}
(\tilde\sigma_{\mu\nu})^{\dot\alpha}{}_{\dot\beta}
\nabla_{\alpha}{}^{\dot\beta} H^{\alpha}\big], \cr}
\eqn\newdos
$$
and,
$$
\eqalign{
T^\laga_0 = &
H^\alpha\nabla_\gamma{}^{\dot\beta}
\nabla_{\alpha\dot\beta}\overline H^\gamma
+\overline H^\alpha \nabla_{\alpha\dot\beta} \nabla_\gamma{}^{\dot\beta}
H^\gamma +i\overline v^{\dot\beta}\nabla_{\alpha\dot\beta} u^\alpha
-i v^{\dot\beta}\nabla_{\alpha\dot\beta}\overline u^\alpha \cr
& -  i \overline u^\alpha \nabla_{\alpha\dot\beta} v^{\dot\beta}
+ i u^\alpha \nabla_{\alpha\dot\beta} \overline v^{\dot\beta}
-2\overline K^\alpha K_\alpha. \cr}
\eqn\newtres
$$
In eq. \newdos\ $\sigma_\mu = ({\bf 1},{\sigma_i})$, where ${\sigma_i}$,
$i=1,2,3$, are the Pauli matrices, and $\sigma_{\mu\nu}$
($\tilde\sigma_{\mu\nu}$) are the generators of Lorentz transformations on
undotted (dotted) spinors.
Although $T^\laga$ is not $Q$-exact it has the property that it vanishes
on-shell. In theories where the action is $Q$-exact, as it is the case
here, this lack of $Q$-exactness of the energy momentun-tensor does not
break the topological symmetry.
Standard arguments  [\tqft] show that in this case the classical limit is
exact and therefore terms which vanish on-shell are harmless.

The vacuum expectation value of any product of operators which are
invariant under $Q$ leads to topological invariants. Actually, the same
arguments show that the resulting vacuum expectation values are in this
case also invariant under  deformations of the gauge connection $A_\mu$.
The gauge current turns out to be $Q$-exact:
$$
J_{\alpha\dot\alpha} = \big\{ Q, i\overline H^\alpha v^{\dot\beta}
+ i \overline v^{\dot\alpha} H^\alpha \big\}.
\eqn\newseis
$$
Unfortunately, the form of the $Q$-transformations of the fields
in \yago\ indicates that there are not operators invariant under $Q$ in
this theory. One could make, however, the observation that according to
\yago\ the field $u_\alpha$ $Q$-transforms into the auxiliary field
$K_\alpha$. A similar situation to this occurs in type B sigma models in
two dimensions \REF\vafa{C. Vafa\journal\mpl&A6(91)337} \REF\witten{E.
Witten, ``Mirror Manifolds and Topological Field Theory", in {\it Essays
on Mirror Manifolds},  ed. S.-T.Yau (International Press, 1992)}
\REF\vafados{C.
Vafa, ``Topological Mirrors and Quantum Rings",
in {\it Essays on Mirror Manifolds}, ed. S.-T.Yau
(International Press, 1992)} [\vafa,\pablo,\witten,\vafados]. When an
operator is not anihilated by $Q$ but it is proportional to auxiliary
fields it may lead to topological invariants. Let us analyze the
situation in this case.

Let us consider, for example,  the following operator,
$$
\phi(P)^{n} = (\overline{u}^\alpha(P) u_\alpha(P))^n,
\eqn\lauramos
$$
where $P\in M$. From \yago\ it follows that,
$$
[Q,\phi^{n}] = -n (\overline{u}^\gamma u_\gamma)^{n-1}
(\overline{K}^\alpha u_\alpha - \overline{u}^\alpha K_\alpha).
\eqn\anton
$$
The vacuum expectation value of the operator \lauramos\ has the
following dependence on $e^{a\mu}$ and $A^\mu$,
$$
\eqalign{
e^a_\nu(P') {\delta\over e \delta e^{a\mu} (P')}
\langle \phi(P)^{n} \rangle = &
\int [DX] \phi(P)^{n} T^\laga_{\mu\nu}(P')
\exp(-{\cal L}^\laga_0),\cr
{\delta\over\delta A^{\alpha\dot\alpha}(P')}
\langle \phi(P)^{n} \rangle = &
\int [DX] \phi(P)^{n}J_{\alpha\dot\alpha}(P')
\exp(-{\cal L}^\laga_0),\cr}
\eqn\chon
$$
where $[DX]$ denotes the full functional integral measure. Using
\newuno, \newseis, the fact that the action es $Q$-exact, and assuming
that $[DX]$ is invariant under $Q$, it turns out that if $P\neq P'$:
$$
{\delta\over\delta e^{a\mu} (P')}
\langle \phi(P)^{n} \rangle =  0,
\,\,\,\,\,\,\,\,\,\,\,\,\,\,
{\delta\over\delta A^{\alpha\dot\alpha}(P')}
\langle \phi(P)^{n} \rangle =  0.
\eqn\joaquina
$$
To get \joaquina\ one just has to realize that the quadratic terms in
the auxiliary fields multiplying the exponential of the action do not
occur at  coincident points for $P\neq P'$. On the other hand, it also
holds that the vacuum expectation value $\langle \phi(P)^{n} \rangle $
is independent of the point $P$. This follows from the fact that
$d\phi^n$ is $Q$-exact up to terms linear in the auxiliary fields:
$$
d\phi^n = n\big\{ Q, (\overline{H}^\alpha d u_\alpha -
d\overline{u}^\alpha  H_\alpha)(\overline{u}^\gamma u_\gamma)^{n-1}
\big\}
+ ({\hbox{\rm linear}}\,\,\, K {\hbox{\rm -terms}}).
\eqn\lara
$$
Thus, the operators $\phi^n$ lead to
quantities which are invariant under small deformations of the vierbein
and the gauge connection.

These arguments show that in order to build non-trivial observables one
must study the cohomology of $Q$ modulo terms linear in the auxiliary
fields $K_\alpha$ and $\overline K_\alpha$. The form of the
transformations \castana\ indicates that this cohomology is trivial
unless one could regard quantities like $\overline H^\alpha H_\alpha$ as
a local coordinate on some manifold with non-trivial topology. We will
not analyze that possibility in this paper.

Due to the quadratic form
of the action ${\cal L}^\laga_0$ and the presence of the symmetry $Q$,
the computation of the vacuum expectation value of products of
$Q$-invariant operators modulo linear $K$-terms reduces to an
integration over zero modes. In other words, after expanding the fields
entering the functional integral into zero and non-zero modes, the
integration of the last ones reduces to a ratio of determinants whose
value is 1. The presence of zero modes leads to a ghost number anomaly
which as usual implies certain selection rule for the observables of the
theory. Let us study the form that this selection rule takes in a
situation which will be of interest in the analysis of the full theory.
Consider the case in
which $M$ is $S^4$ and $A_\mu$ is an $SU(2)$ anti-self-dual connection
($F^+_{\alpha\beta}=0$) of second Chern class $k$. In this case, since
$R>0$, there are not $H_\alpha$-zero modes in  ${\cal L}^\laga_0$.
Respect to the spinor fields $u_\alpha$ and $v_{\dot\alpha}$,
their structure of zero modes depends on the representation chosen.
If they belong to the $SU(2)$ representation of isospin $j/2$, there
are ${1\over 6}j(j+1)(j+2)k$ $v_{\dot\alpha}$-zero modes, while there are
not $u_\alpha$-zero modes \REF\jare{R. Jackiw and C.
Rebbi\journal\pr&D16(77)1052} [\jare]. Thus, the selection rule that
emerges is that operators which could possibly lead to a non-zero vacuum
expectation value would contain ghost number
$-{1\over 6}j(j+1)(j+2)k$. This selection rule by itself is strong enough
to argue that there are not non-trivial observables  in the situation
considered since  the form of the transformations \castana\ indicates
that there are not $Q$-invariant operators of negative ghost number.

The topological matter theory described by the action
${\cal L}^\laga_0$ in \maria\ does not seem to provide  observables
leading to topological invariants. However, this theory can be coupled to
topological Yang-Mills or topological gravity modifying the values of the
observables of those theories. In the rest of this paper we will describe
the topological matter coupling to the first of these theories.

Before describing the coupling, let us recall first the structure of
Donaldson-Witten theory. The action of this
theory is [\tqft], $$
\eqalign{
{\cal L}^{\hbox{\sevenrm DW}} = &{1\over g^2} \int_M d^4x
e \tr\Big[ Q,
\, {1\over 4} \big(F_{\alpha\gamma}^{+}+
G_{\alpha\gamma}\big) \chi^{\alpha\gamma}
+i\lambda\D_{{\alpha\dot\beta}}\psi^{\alpha\dot\beta}
+{i\over2}\lambda\big[\eta,\phi\big]\Big] \cr
= & {1\over g^2} \int_M d^4x e \tr \Big[{{1\over 4}\big(F^{+}\big)}^2
- {1\over 4}{(G)}^2  -
\chi^{\alpha\gamma}\D_{{\alpha\dot\beta}}\psi_{\gamma}^{\,\,\dot\beta}
+{i\over4}\phi\big\{\chi_{\alpha\gamma},\chi^{\alpha\gamma}\big\} \cr &
+\eta\D_{{\alpha\dot\beta}}\psi^{\alpha\dot\beta}
-i\lambda\big\{\psi_{\alpha\dot\beta},
\psi^{\alpha\dot\beta}\big\}
-\lambda\D_{{\alpha\dot\beta}}\D^{{\alpha\dot\beta}}\phi
+{i\over2}\phi\big\{\eta,\eta\big\}+{1\over2}
{\big[\lambda,\phi\big]}^2
\Big], \cr}
\eqn\pala
$$
where $\lambda$ and $\phi$ are commuting scalar fields, and $\eta$,
$\psi_{\alpha\dot\beta}$ and $\chi_{\alpha\beta}$ are anticommuting
scalar, vector and anti-self-dual fields, respectively.
$G_{\alpha\beta}$ is an auxiliary commuting and anti-self-dual field
\REF\action{J.M.F. Labastida and M. Pernici\journal\pl&212B(88)56}
[\action]. In \pala\ $g$ denotes the gauge coupling constant. The
operator $Q$ in \pala\ is the corresponding one to the following
transformations, $$ \eqalign{
\delta A_{{\alpha\dot\beta}} &=\epsilon
\psi_{\alpha\dot\beta}, \cr
\delta \psi_{\alpha\dot\beta} &=-\epsilon
\D_{{\alpha\dot\beta}} \phi, \cr
\delta \lambda &= \epsilon\eta, \cr
\delta \eta &= i\epsilon[\lambda, \phi], \cr}
\qquad\qquad
\eqalign{
\delta \chi_{\alpha\beta} &= \epsilon(F^{+}_{\alpha\beta} -
G_{\alpha\beta}), \cr
\delta G_{\alpha\beta}&=
\epsilon\big(\D_{{(\alpha\dot\gamma}}
\psi_{\beta)}^{\,\,\dot\gamma}
-i\big[\chi_{\alpha\beta}, \phi\big]\big), \cr
\delta \phi &= 0. \cr}
\eqn\cubo
$$
Since $Q^2$ is just a gauge transformation the action \pala\ is
manifestly invariant under the transformations \cubo.

The observables of Donaldson-Witten theory, which lead to topological
invariants, are arbitrary products of operators [\tqft],
$$
{\cal O}^{(\gamma)}=\int_\gamma W_{k_\gamma},
\eqn\observables
$$
where $\gamma$ is a homology cycle of $M$ of dimension $k_\gamma$, and
$W_{k_\gamma}$ is one of the differential forms,
$$
\eqalign{
W_0 = & {1\over 2} \tr \phi^2, \cr
W_1 = & \tr(\phi\wedge\psi), \cr
W_2 = & \tr({1\over 2}\psi\wedge\psi + i \phi\wedge F), \cr
W_3 = & i\tr(\psi \wedge F),\cr
W_4 = & - {1\over 2} \tr (F\wedge F). \cr}
\eqn\obs
$$

The coupling of the topological matter described in \maria\ to the
full topological multiplet of Donaldson-Witten theory can be obtained
by twisting its corresponding $N=2$ counterpart. It turns out that the
resulting theory can be truncated making it simpler. In what follows we
will describe the truncated theory. Details of the truncation will be
presented elsewhere. The full  action takes the following form,
$$
{\cal L} = {\cal L}^{\hbox{\sevenrm DW}} + {\cal L}_0 + m{\cal L}_m,
\eqn\playa
$$
where,
$$
\eqalign{
{\cal L}_0
= & \int d^4x e  \Big[
\overline{H}^{\alpha}(\tableau{1}+{1\over 4}R) H_{\alpha}
+{i\over 2}\overline{H}^\alpha F_{\alpha\beta}^{+} H^\beta +{i}
\overline{u}^{\alpha}\nabla_{{\alpha\dot\alpha}}{v}^{\dot\alpha} -
{i}\overline{v}^{\dot\alpha}\nabla_{{\alpha\dot\alpha}}u^{\alpha}
\cr
&\,\,\,\,\,\,\,\,\,\,\,\,\,\,\,\,\,\,\,\,\,\,\,\,\,\,\,\,\,\,
+\overline{K'}^{\alpha}K_{\alpha}'
+\overline{H}^{\alpha}\psi_{\alpha\dot\beta}v^{\dot\beta}
-\overline{v}^{\dot\beta}\psi_{\alpha\dot\beta}
H^{\alpha}+i\overline v^{\dot\alpha}\phi v_{\dot\alpha}\Big],\cr
 m {\cal L}_m = &
 \int d^4x e \big[m^2 \overline{H}^\alpha H_\alpha
+m(\overline{u}^\alpha u_\alpha +
\overline{v}^{\dot\alpha}v_{\dot\alpha})
-i m\overline{H}^{\alpha}\phi H_{\alpha} \big].\cr}
\eqn\rastri
$$

The coupling modifies the  $Q$ and $Z$-transformations \yago.
They take now the following form:
$$
\eqalign{
\delta H_{\alpha} =& \epsilon u_{\alpha}, \cr
\delta u_{\alpha} =& -\epsilon ( K_{\alpha} +i\phi H_{\alpha}), \cr
\delta v_{\dot\alpha} =&
i\epsilon\D_{{\alpha\dot\alpha}} H^{\alpha}, \cr
\delta K_{\alpha} =&i\epsilon\nabla_{{\alpha\dot\alpha}}
v^{\dot\alpha},  \cr}
\qquad\qquad
\eqalign{
\delta_z {H}_{\alpha} =&  -z(K_{\alpha}+i\phi H_\alpha), \cr
\delta_z {u}_{\alpha} =&  -{i}z(
\D_{\alpha\dot\alpha}v^{\dot\alpha}+\phi u_\alpha),\cr
\delta_z {v}_{\dot\alpha} =&z(
i\D_{\alpha\dot\alpha}u^{\alpha}+\psi_{\alpha\dot\beta}H^\alpha),\cr
\delta_z {K}_{\alpha} =&z
\big[(\tableau{1}+{1\over 4}R) H_\alpha
+{i\over 2} F^{+}_{\alpha\beta}H^\beta +
\psi_{\alpha\dot\alpha}v^{\dot\alpha}\big].\cr}
\eqn\trillo
$$
Of course, one has a
similar set of transformations for the fields with overlines. In the
coupled theory the algebra $2Q^2=Z$ holds. The $z$-transformations of
the fields in the topological Yang-Mills multiplet are just gauge
transformations with gauge parameter $z\phi$.

The action \playa\ represents the coupling of topological matter to
Donaldson-Witten theory. Let us analyze the structure of the resulting
theory. Certainly, if $m\neq 0$ the topological character of the
theory is broken. For $m=0$ it turns out that contrary to the case  of
${\cal L}^\laga_0$ in \maria\ the action ${\cal L}_0$ in \rastri\ is
not $Q$-exact. This can be demonstrated writing the most general local
expression of ghost number $-1$ (mod 4) and showing that it does not
lead to ${\cal L}_0$. The $Z$-symmetry of the theory is very
restrictive and somehow the responsible for the non-existence of a
reasonable matter action which is $Q$-exact for the coupled theory.
This has very important consequences. As shown below, the energy-momentun
tensor of the theory is not $Q$-exact. As in the non-coupled case, the
part of the energy-momentun tensor which is not $Q$-exact vanishes
on-shell. However, in the coupled case, since the action is not
$Q$-exact, one does not possesses an argument to disregard the terms
in the energy-momentum tensor which are not $Q$-exact and therefore the
manifiest topological character of the theory is broken. Matter couplings
to Donaldson-Witten theory lead in principle to non-topological
corrections to Donaldson invariants, \ie, to the vacuum expectation
values of the observables \observables\ computed with ${\cal
L}^{\hbox{\sevenrm DW}}+{\cal L}_0$.

Let us analyze the computation of the vacuum expectation value of a
product of observables of the type \observables\ of ghost number $r$.
We will consider\foot{We make this
choice to be specific but a similar discussion holds in general.}
a situation in which $M$ is $S^4$ and the gauge group
is $SU(2)$, while the matter fields are valued in a $SU(2)$
representation of dimension $d$. Since ${\cal L}^{\hbox{\sevenrm DW}}$
is $Q$-exact, standard arguments show that the vacuum expectation
values are invariant under deformations of the parameter $g$ in \pala.
This allows to make computations of $Q$-invariant quantities in the
limit $g\rightarrow 0$, in which the contributions from the functional
integral are dominated by the classical configurations of
Donaldson-Witten theory.

The classical configurations of the gauge connection are anti-self-dual
gauge fields ($F_{\alpha\beta}^+=0$). For the case in
which $M$ is $S^4$ and the gauge group is $SU(2)$ these form a moduli
space of dimension $8k-3$, where $k$ is the second Chern-class. There
are in addition $8k-3$ $\psi_\mu$-zero modes which are anticommuting.
There are not  zero modes for the rest of the fields in the gauge
multiplet. The contribution from matter fields is computed spanding
them around classical configurations.  For an anti-self-dual gauge
connection of second Chern class $k$, and matter fields in the $SU(2)$
representation of isospin $j/2$, there are ${1\over 6}j(j+1)(j+2)k$
$v_{\dot\alpha}$-zero modes. Since $v_{\dot\alpha}$ has  ghost number
$-1$, in order to have a non-vanishing vacuum expectation value the ghost
number of the operator entering the functional integral must take the
value $r=8k-3-{1\over 6}j(j+1)(j+2)k$.
The computation of observables can be carried out exactly using
the invariance under variations of $g$. The resulting expression
involves, besides integrations of zero modes, convolutions of these
with fermionic and bosonic propagators.

The energy-momentum tensor $T_{\mu\nu}$ corresponding to
${\cal L}^{\hbox{\sevenrm DW}}+{\cal L}_0$ takes the form:
$$
T_{\mu\nu}  =
\big\{ Q ,\Lambda_{\mu\nu}+\Lambda_{\mu\nu}^\laga \big\}
+ {1\over 2}g_{\mu\nu}  T^\laga,
\eqn\temunuzero
$$
where $\Lambda_{\mu\nu}$ corresponds to the part of topological
Yang-Mills [\tqft], $\Lambda_{\mu\nu}^\laga$ is the one given in \newdos,
and, $$
\eqalign{
T^\laga = &
H^\alpha\nabla_\gamma{}^{\dot\beta}\nabla_{\alpha\dot\beta}\overline
H^\gamma +\overline H^\alpha \nabla_{\alpha\dot\beta}
\nabla_\gamma{}^{\dot\beta} H^\gamma +i \overline
v^{\dot\beta}\nabla_{\alpha\dot\beta} u^\alpha -i
v^{\dot\beta}\nabla_{\alpha\dot\beta}\overline u^\alpha
 -  i \overline u^\alpha \nabla_{\alpha\dot\beta} v^{\dot\beta}\cr
& + i u^\alpha \nabla_{\alpha\dot\beta} \overline v^{\dot\beta}
-2\overline K^\alpha K_\alpha
-2 \overline H^\alpha \psi_{\alpha\dot\beta} v^{\dot\beta}
+2 \overline v^{\dot\beta} \psi_{\alpha\dot\beta}  H^\alpha -2 i
\overline v^{\dot\alpha}\phi v_{\dot\alpha}. \cr}
\eqn\newcuatro
$$
This last quantity vanishes on-shell. However, since the action of the
theory is not $Q$-exact one can not ignore those terms when computing
the dependence of vacuum expectation values on the vierbein. For an
arbitrary variation of the vierbein one finds,
$$
{\delta\over \delta e^{a\mu} }
\langle \prod {\cal O}^{(\gamma)} \rangle =\half e e_{a\mu} \langle
\prod{\cal O}^{(\gamma)} T^\laga \rangle
\eqn\lae
$$
where $\prod {\cal O}^{(\gamma)}$ denotes an arbitray product of the
operators \observables.

Equation \lae\ indicates that when coupling topological matter to
topological Yang-Mills the manifiest topological character of the theory
is lost. It is important to study the properties of the vacuum
expectation value in the right hand side of \lae. For example, it would
be interesting to  characterize the topologies for which it vanishes.
If the right hand side
of eq. \lae\ vanishes this theory leads to a set of invariants which are
richer than Donaldson invariants since they are also labeled by the
$SU(2)$ representation carried by the matter fields.

Donaldson invariants are polynomials on
$H_*(M)\times H_*(M)\times ...\times H_*(M)$, \ie,
products of the homology groups of $M$. The vacuum expectation
values of the observables \observables\ evaluated in the coupled
theory presented in this paper are also polynomials on $H_*(M)\times
H_*(M)\times ...\times H_*(M)$. Since the action is $Q$-invariant the
quantities   $\langle \prod {\cal O}^{(\gamma)} \rangle$ are invariant
under deformations of the cycles $\gamma$. The argument is the standard
one. Since a deformation of $\gamma$ leads to a $Q$-exact deformation of
${\cal O}^{(\gamma)}$, the corresponding vacuum expectation value
vanishes due to the $Q$-invariance of the action. This argument holds
also in the case $m\neq 0$, so even in this case the observables can be
regarded as polynomials on $H_*(M)\times H_*(M)\times ...\times H_*(M)$.

The observables computed with the action
${\cal L}^{\hbox{\sevenrm DW}}+{\cal L}_0$ are polynomials on
$H_*(M)\times
H_*(M)\times ...\times H_*(M)$ which
depend on the representation of $SU(2)$ which have been chosen for
the matter fields. The resulting quantities may not be topological
invariants but still be interesting quantities. Whether or not these
quantities may help in the study of four-dimensional spin manifolds is an
open question. If the
observables were computed with ${\cal L}^{\hbox{\sevenrm DW}}+{\cal
L}_0 +m{\cal L}_m$ a dependence on $m$ would be introduced and the
resulting  equation \lae\ would possess additional terms.

The breaking of the topological symmetry which may occur in this theory
would imply that the theory possess propagating modes. The breaking is
caused by the interaction since in the absence of coupling one is left
with two theories which are topological. Many
questions should be answered from this point of view. For example, one
would like to know if the degeneracy of the vacuum  of Donaldson-Witten
theory is lifted by the interaction, or if the breaking leads to the
presence of physical degrees of freedom.

In a forthcoming paper we will study the theory presented in this
letter in full detail. We will analyze  its
symmetries and its features from both a physical and a mathematical
point of view. Also, it would be very interesting to construct
different types of topological matter based on other $N=2$
multiplets like, for example, the relaxed hypermultiplet
\REF\relaxed{P.S. Howe, K.S. Stelle and P.K.
Townsend\journal\np&B214(83)519} [\relaxed].

\ack
We would like to thank A.V. Ramallo and J. Mas for very helpful
discussions. This work was supported in part by DGICYT under grant
PB90-0772, and by CICYT under grants AEN88-0013 and AEN88-0040.

\refout


\end